# Spin-active chlorine-related centers in 4H-SiC with telecom-band emissions

Danial Shafizadeh,[1] Misagh Ghezellou,[1] Viktor M. Bobal,[2] Lasse Vines,[2] Jawad Ul-Hassan,[1] Valdas Jokubavicius,[1,3] Nguyen T. Son,[1,*] and Ivan G. Ivanov[1]

[1] *Department of Physics, Chemistry and Biology, Linköping University, SE-58183, Linköping, Sweden*

[2] *Department of Physics, Centre for Materials Science and Nanotechnology, University of Oslo, 0316 Oslo, Norway*

[3] *TekSiC, Molijns Väg 3, SE-58941 Linköping, Sweden*

A photoluminescence (PL) and magnetic resonance investigation of a defect in chlorine-implanted 4H-SiC is presented. This Cl-related center emits light at telecom wavelengths with zero-phonon lines in the range 1350 – 1540 nm. Its four configurations exhibit stable PL spectra characterized by narrow zero-phonon lines. For the two configurations that emit light at the C-band, a Debye–Waller factor in the range 22-25% is estimated. Optically detected magnetic resonance confirms that the Cl-related center is spin active and stable at room temperature with the zero-field splitting in the range of 1.0–1.4 GHz. The combined optical and spin properties suggest this center to be a highly promising candidate for scalable quantum networks.

Spin-active color centers in semiconductors, such as diamond, present compelling advantages for quantum information processing. These benefits include optically addressable spins with long coherence times under ambient conditions, access to low-abundance nuclear spins for quantum memory applications, and high-fidelity spin-to-photon interfaces. Among these centers, the nitrogen-vacancy in diamond (NV) – developed over decades – remains the benchmark for performance with recent demonstration of deterministic entanglement and teleportation across a three-node network [1]. Entanglement of two nuclear spins of the Si-vacancy center (SiV) in diamond over a 35-km long fiber loop has also been demonstrated recently [2]. For applications in quantum networks, frequency conversion of their visible emissions to telecom wavelengths is necessary. However, this process is hindered by low conversion efficiency – approximately ~17% for the NV center [3] and ~30% for the SiV center [2]. These limitations motivate the search for more suitable color centers in alternative material platforms, particularly silicon carbide (SiC), which shares most of the advantages of wide-bandgap diamond while offering a wafer-scale material with established integrated circuit technologies and nanofabrication methods – critical enablers for scalable quantum applications [4,5]. Several color centers in SiC emit light in the near-infrared regions [6], including the neutral vanadium center ($V^{4+}$), which exhibits emission at the telecom O-band [7]. Coherent control [8] and ultra-narrow inhomogeneous spectral distribution [9] of V qubits have been demonstrated. However, the requirement of sub-Kelvin temperatures to extend spin relaxation times [10] remains a drawback. The $Er^{3+}$ impurity center in various semiconductors, including SiC [11] and Si [12], emits in the C-band (1540 nm). Silicon also hosts intrinsic defects, such as the G [13] and T [14] centers, which emit in the telecom O-band. However, both the $Er^{3+}$ and T centers exhibit weak emission due to their long radiative lifetimes, while the G center has an undesirable electronic structure, with a singlet ground state and a metastable spin-triplet state [15]. These unfavorable optical and spin properties make these centers less attractive for quantum-communication applications.

Recent density functional theory (DFT) calculations predict a chlorine-vacancy defect in 4H-SiC, consisting of a chlorine atom substituting a carbon site and a neighboring silicon vacancy in the positive charge state ($ClV^+$). The defect is predicted to be NV-like with spin S = 1 and emit light at telecom wavelengths [16]. This theoretical work has inspired the present study.

In this Letter, we report our realization of a Cl-related spin-active defect in 4H-SiC which exhibits emissions in the telecom bands. Photoluminescence (PL) measurements suggest that the defect possesses four distinct configurations with narrow zero phonon lines (ZPLs) at 1539.2, 1533.3, 1397.0, and 1351.0 nm, denoted Cl1–Cl4, respectively. The PL emission remains stable under the excitations used in this work. The C-band centers Cl1 and Cl2 show rather sharp phonon replicas due to local vibrational modes and negligible phonon sidebands with an estimated Debye-Waller (DW) factor in the range 22-25%. Optically detected magnetic resonance (ODMR) confirms that the defects are spin-active, with zero-field splitting (ZFS) ranging from 1.0 to 1.4 GHz.

During preparation of this manuscript, we encountered another PL study of a Cl-related defect in 4H-SiC by Anisimov and co-workers [17] and a revised DFT calculation of ZFS for the ClV defect [18] appearing on arXiv. Although some of the PL lines in [17] are apparently the same as ours, the overall PL results and their interpretation are markedly different from ours. Furthermore, although the revised ZFS values in [18] are close to our experimental values, it is unlikely that our Cl-related defect has the predicted spin S=1 of the ClV⁺ center. Our findings indicate that the observed Cl-related defect, owing to its favorable optical and spin properties, holds strong promise for enabling scalable quantum networks.

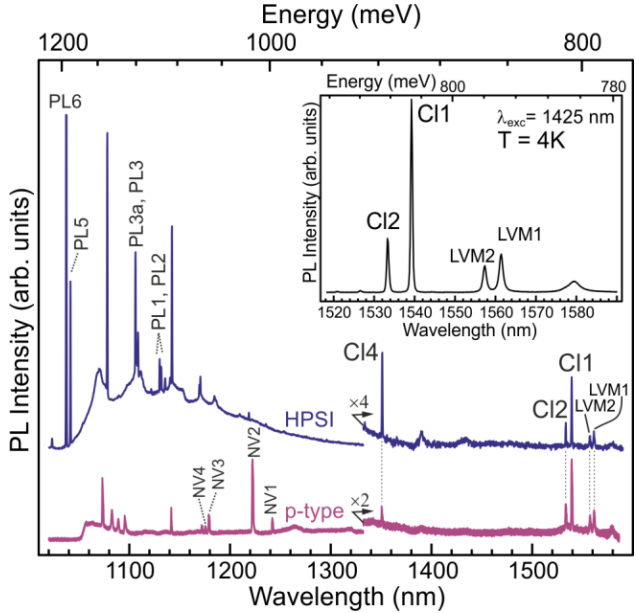

Fig. 1. Low-temperature PL spectra of Cl-implanted 4H-SiC: (top) HPSI 4H-SiC substrate, and (bottom) p-type layer. The excitation is 990 nm and the emissions are collected along the crystallographic c-axis, hence the Cl3 line does not appear (see text). The ZPLs at wavelengths shorter than 1300 nm belong to the divacancy and other impurities (see description in text). The inset is a high-resolution spectrum illustrating the difference in the ZPL linewidths of Cl1 and Cl2 (~0.3 meV close to the spectral resolution) and the LVM replicas LM1 and LVM2 (~0.5 meV). Notice the scale changes in the spectra.

The samples used in this work are high-purity semi-insulating (HPSI) 4H-SiC substrates and p-type epitaxial 4H-SiC layers (~25 μm thick) grown by chemical vapor deposition (CVD) doped with Al to a concentration of ~$5\times10^{14}$ cm$^{-3}$. The samples are implanted with Cl⁺ ions with different energies ranging from 36 keV to 2 MeV to form an implanted box with the Cl concentration in the range of ~$1\times10^{16}$ cm$^{-3}$ (details of SRIM simulations are described in the Supplementary Information [19]). During implantations, the samples are kept at ~600 °C to reduce the formation of interstitial-related complexes and clusters. Implanted samples are annealed for one hour at 1600 or 1800 °C in Ar ambient.

The PL emission is dispersed by a Jobin-Yvon HR460 monochromator using a 300 grooves/mm grating and detected by an InGaAs multichannel detector. Tunable Ti:sapphire laser or diode-based Toptica lasers are used for optical excitation, enabling selective excitation over a wide range. The sample is mounted in a Montana S50 closed-cycle cryostat equipped with microscope objective (Olympus LCPLAN N 50X IR), allowing temperature regulation from 3.5 K to room temperature. For ODMR measurements, a 50 μm-diameter copper wire was positioned at the sample surface, hence, oriented perpendicular to the c-axis of the crystal, to serve as a microwave antenna. After filtering the laser, the integrated PL emission is coupled to a superconducting nanowire single photon detector (SNSPD) from Quantum Opus connected to a time-correlated single-photon counting (TCSPC) module (HydraHarp 400, PicoQuant).

The PL spectra of Cl-implanted p-type layer and HPSI substrate in the spectral range of 1000 – 1600 nm measured at 4 K under 990 nm excitation are shown in Fig. 1. Some of the lines in the spectral range below 1300 nm in the HPSI samples are identified as divacancy-related lines (PL1 – PL6 [5], and PL3a [20]), and in the p-type layer as weak nitrogen-vacancy emission (NV1 – NV4) [21,22]. The rest of the lines below 1300 nm belong to unidentified defects. In both p-type and HPSI samples, the spectral region above 1300 nm shows the lines associated with the Cl-related center, and these are the only lines common for both samples. With PL detected along the crystal c-axis we detect only luminescence polarized perpendicular to the crystal c-axis (E⊥c). In this configuration the Cl-related emission shows a ZPL at 1351.0 nm, two sharp lines in the C-band (1533.3 and 1539.2 nm), and two broader lines at 1557.2 and 1561.2 nm (denoted LVM2 and LVM1 in Fig. 1, respectively). Initially, we associate the latter two lines with local-phonon replicas of the ZPLs Cl2 and Cl1 mainly because of their broader linewidth compared to the ZPLs (0.5 vs 0.3 meV, cf. the inset of Fig. 1), however, further evidence corroborating this notion is presented in the discussion of the temperature dependence later. From the energy separation between Cl2 and LVM2 and Cl1 and LVM1 we obtain the local-phonon energies 12.4 and 11.5 meV, respectively. Furthermore, the broader band at ~1579 nm is separated by 22.5 meV from Cl2 and 20.5 meV from Cl1 and may be composed of the overtones of the above LVMs (cf. the inset in Fig. 1), in analogy to the PL of Mo [25].

It is known that the different configurations of a certain defect may be sensitive to the polarization of the excitation laser and may have different PL-polarization properties [20,23]. For instance, the emission of the V2 configuration of the Si vacancy in 6H-SiC has $E_{PL}\|c$ polarization and can only be detected in a direction perpendicular to the c-axis (i.e., from the edge of a sample with the common (0001) orientation), while the PL emissions of the V3 configuration has E⊥c polarization and can be detected both from the surface and the edge of an (0001)-oriented sample [23]. The excitation-laser polarization is also important. Thus, the axial PL1 and PL2 configurations of the divacancy in 4H-SiC cannot be excited with laser polarization $E_L||c$ [20,24]. Therefore, we examine the polarization properties of the PL using different polarization of the excitation laser with respect to the c-axis, $E_L||c$ and $E_L⊥c$. For these measurements the sample is mounted edge-on under the objective. To avoid interference from emissions of the divacancy and other defects at shorter wavelengths, we measure PL in the range 1300-1600 nm using 1290 nm excitation.

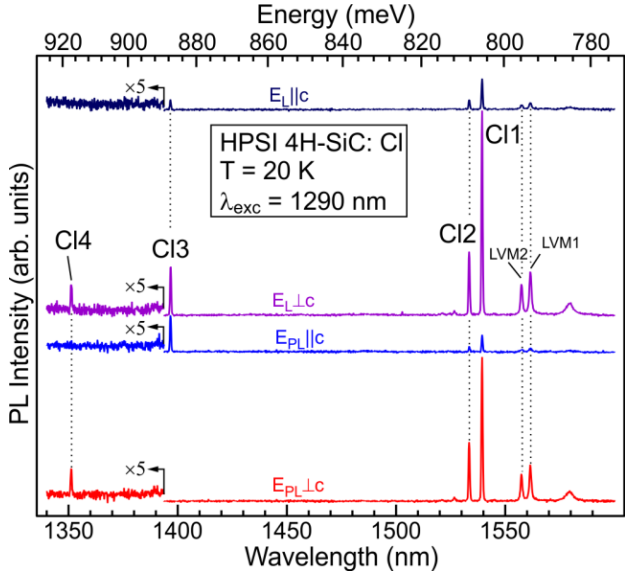

Fig. 2. Low-temperature PL spectra of the chlorine-related defects measured at 20 K under 1290 nm excitation with different laser polarizations with respect to the c-axes (top two curves) and different PL polarizations (bottom two curves). Note the scale changes below ~ 1395 nm.

Fig. 2 displays the polarization properties measured on the HPSI sample at 20 K. The top two spectra show the excitation efficiency with two different laser polarizations ($E_L||c$ and $E_L⊥c$) and the bottom two spectra show the PL polarization excited with $E_L⊥c$. In the top two spectra, both PL with $E_L||c$ and $E_L⊥c$ are detected. However, the spectrum with $E_L||c$ is much weaker than that with $E_L⊥c$ excitation and is probably excited from $E_L⊥c$ polarization "leaking" through the surface of the sample, which can be understood as follows. Since the implanted active region of the sample is very close to the surface (~ 500 nm), when we measure with $E_L||c$ we cannot exclude a minute part of the excitation laser penetrating the sample through the sample surface due to the extreme proximity of the laser spot to the surface of the sample. This latter "leak" through the sample surface enables weakly the undesired $E_L⊥c$ excitation which may explain the weak appearance of the spectrum with $E_L||c$. Therefore, this consideration suggests that most likely all the ZPLs (Cl1-Cl4) cannot be excited by laser polarization $E_L||c$. On the other hand, the PL spectrum obtained with $E_L⊥c$ (second from top in Fig. 2) shows four ZPLs lines labeled Cl1-Cl4, at 1539.2, 1533.3, 1397.0, and 1351.0 nm, respectively. We then examine the polarization of these lines using a polarizer in the PL path (bottom two spectra in Fig. 2). We find that the Cl1, Cl2, and the Cl4 lines have $E_{PL}⊥c$ polarization (bottom curve) while the Cl3 ZPL is polarized $E_{PL}||c$. We notice that some emission with $E_{PL}⊥c$ can still leak from the surface of the sample when the spectrum with $E_{PL}||c$ is measured for the same reasons as discussed above in relation to the experiment with $E_L||c$. This explains the weak appearance of Cl1 and Cl2 lines in the second from bottom curve of Fig. 2 ($E_{PL}||c$). However, when the polarizer is set to measure $E_{PL}⊥c$, the PL emission with $E_{PL}||c$ is completely suppressed, hence, no contribution from the Cl3 line polarized $E_{PL}||c$ is observed in the bottom spectrum of Fig. 2. Thus, the emissions of Cl1, Cl2, and Cl4 have polarization $E_{PL}⊥c$ while that of Cl3 has polarization $E_{PL}||c$. The phonon replicas LVM1 and LVM2 also show distinct $E_{PL}⊥c$ polarization.

Next, we examine the temperature dependence of the Cl1 – Cl2 related PL spectrum in the range 3.5 – 120 K (Fig. 3). We notice that the Cl1 and Cl2 retain their intensity ratio. In our experiments, we cannot entirely eliminate the possibility that Cl1 and Cl2 belong to a single defect configuration with split ground state, as stated in [17]. However, the two lines denoted LVM1 and LVM2 at 1557.2 and 1561.2 nm which also show no thermalization, are not due to another configuration with split ground state as suggested in [17]. As already discussed, we assign these two lines to phonon replicas involving local phonons of the Cl1 and the Cl2 centers. Further evidence confirming this assignment comes from the temperature dependence. At temperatures above 25 K, PL lines emerging from second excited states of Cl1 and Cl2 appear at higher energies, denoted Cl1′ and Cl2′ in Fig. 3. Both lines are separated by ~6.7 meV from the corresponding Cl1 and Cl2 lines. At 50 – 60 K the Cl1′ and Cl2′ become strong enough so that even their LVM-replicas are detected denoted LVM1′ and LVM2′, respectively (inset in Fig. 3). Naturally, the energy separations of LVM1′ and LVM2′ from Cl1′ and Cl2′ mimic exactly the separations of LVM1 and LVM2 from Cl1 and Cl2 (12.4 and 11.5 meV, respectively), which corroborates our assignment of LVM1, LVM2, LVM1′ and LVM2′ to phonon replicas due to local phonons. The line positions and polarization properties of ZPLs, their thermally activated ESs, and the LVMs are given in Table I.

Estimation of the DW factor is complicated for Cl3 and Cl4 due to overlapping emissions at longer wavelengths but possible for the Cl1 and Cl2 centers. Using excitation at 1425 nm, we estimate a DW factor for Cl1 and Cl2 in the range of 22-25% all emissions at wavelengths λ>1500 nm are detected. Details on measurements and analysis are given in Supplementary Information [14].

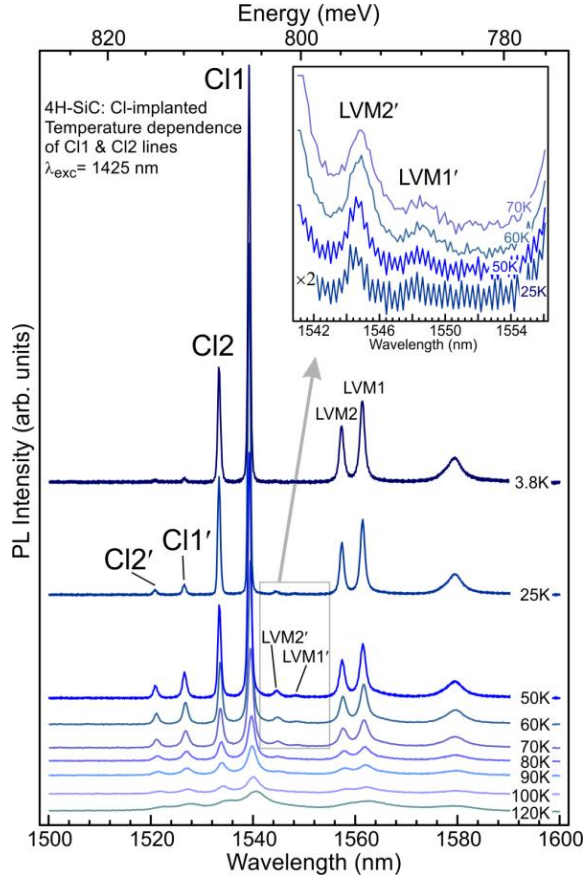

Fig. 3. Temperature dependence of the PL spectra of CL1 and CL2 lines in the p-type sample using 1425 nm excitation optimal for Cl1 and Cl2. With increasing temperature, two new lines Cl1′ and Cl2′ appear separated by ~6.7 meV from Cl1 and Cl2, respectively, indicating the presence of second excited states at ~6.7 meV above the lowest ones for Cl1 and Cl2. The inset zooms on the LVMs of Cl1′ and Cl2′, well distinguishable in the range 25 – 70 K, which mimic the LVMs of Cl1 and Cl2.

Considering the usual deviation in energy of ZPLs between DFT calculations and experiments, e.g., ~150-170 meV for the divacancy in SiC [26], the wavelengths of ZPLs Cl1-Cl4 are in the same spectral region as predicted by DFT for the ClV⁺ defect [16]. However, there are several discrepancies between our experimental data and the theoretically predicted ClV⁺ defect. Firstly, all configurations of the ClV⁺ defect are predicted to have strong phonon sidebands [16], while the phonon sidebands of Cl1 and Cl2 are characterized by sharp LVMs. Moreover, the predicted DW factors for the ClV⁺ defect (0.95-2.88% [16]) are in contrast with the measured values for Cl1 and Cl2 (22-25%). These discrepancies suggest that the Cl-related defect observed experimentally may be different from the theoretically considered one (ClV⁺), e.g., it might be a different charge state of ClV. This notion is corroborated by the ODMR results discussed further below.

We compare now our results with the PL spectra of ClV in [17]. Apparently, our CL4 line at 1351 nm is the same as ClV1 of [17] (1350 nm vs 1351 nm in our study), and the group of lines in the C-band are the same within the experimental accuracy of the two experiments. However, Cl3 is missing in [17] since it has polarization (E$_{PL}$||c) and cannot be detected from the surface of the sample (along the c-direction), which is the only configuration used in [17]. In addition, we believe that the very weak ClV3 line in [17] is not related to this Cl-related center since it is not observed in our spectra. Furthermore, as discussed above the two lines denoted ClV4 in [17] are not ZPLs but the first-order LVMs of Cl1 and Cl2. Finally, the higher-energy peak in the doublet ClV3 does not arise from splitting of the excited state, as demonstrated by the temperature dependence, suggesting that the assignment of the C-band lines in [17] is incorrect.

To examine the spin properties of the Cl-related defects, ODMR measurements were performed in the frequency range from 1000 to 2000 MHz to cover the frequency range predicted by theory for the zero-field splitting (ZFS) of the ClV defects [16]. Fig. 4(a) shows the ODMR spectrum measured under 1290 nm excitation with emissions being detected from the edge of the sample to detect PL from all Cl1-Cl4 configurations. We find signal only in the range ~ 1000 – 1400 MHz. The dependence of the ODMR spectrum on an external magnetic field in the range of 0-20 G is shown as color-coded map in Fig. 4(b). To find a correlation between the PL and the ODMR centers, we repeat the ODMR experiments under 1425 nm excitation and find that the peak at ~1300 MHz in Fig. 4(a) vanishes [Fig. 4(c)]. This result indicates that the ODMR peak at ~1300 MHz may be related to either Cl3 or Cl4 centers with a higher probability that it is from Cl3 since under 1290 nm excitation, the Cl4 configuration is not efficiently excited as one can see comparing its intensity in Fig. 2 (E$_⊥$⊥c) and in Fig. 1 (upper spectrum) measured under 990 nm excitation. It is possible that the ODMR of Cl4 is not detected at all in our experiments. We reduce further the photon energy of the excitation by using a 1525 nm laser. At 4 K, a strong ODMR line at ~1101 MHz and a weak shoulder at 1147 MHz are detected as shown in Fig. 4(d). Note that the ODMR contrast under 1525 nm excitation increases by an order of magnitude compared to that measured under 1290 nm excitation, likely due to reduced background emissions. The ODMR peaks with 1525-nm excitation can be followed all the way to room temperature, at which the shoulder at ~1147 MHz is clearly detected. These two ODMR lines are likely related to the Cl1 and Cl2 configurations since the other configurations are not effectively excited with 1525-nm laser. The appearance of other weak ODMR peaks at ~1197 and ~1255 MHz at room temperature may be due to phonon-assisted excitation of Cl1′ and Cl2′ (the laser is between these two lines), hence, these weak lines may be related to the excited states Cl1′ and Cl2′. Thus, the four ODMR peaks observed at room temperature in Fig. 4(d) are likely related to the Cl1 and Cl2 configurations but individual assignment to Cl1 or Cl2 is not possible in this ensemble study.

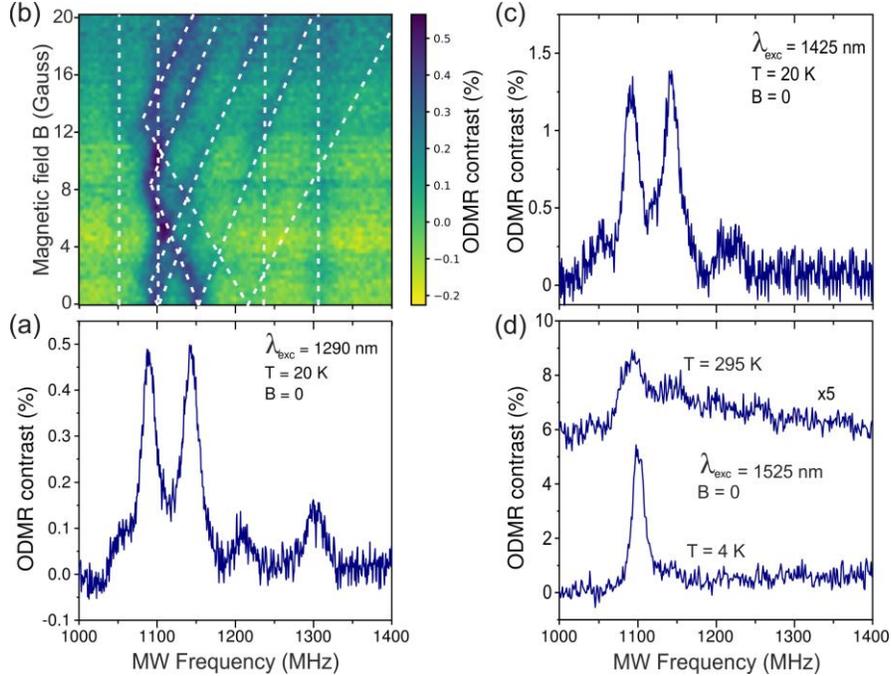

Fig. 4. (a, c, d) Zero-field ODMR spectra of Cl-implanted HPSI 4H-SiC layer measured under different excitations. (a) Four peaks and a weak shoulder (left) are detected under 1290 nm excitation. (b) Dependence of the ODMR spectrum on external magnetic field parallel to the c-axis in the range 0–20 G measured at 4K under 1290 nm excitation. (c) The peak around 1300 MHz seen in part (a) vanishes under 1425 excitation. (d) Under 1525 nm excitation, only one prominent ODMR peak at ~1100 MHz is observed. The measurement temperature is denoted for each of the spectra in (a, c, d). The output MW power is ~30 mW but much lower at the sample [14]. Notice the slight frequency downshift by ~3.5 MHz when the temperature is raised from 4K to 20 K between the peaks in the spectra in (a) and (b), and (c) and (d).

We notice that in the revised DFT calculations [18], the ZFSs are reduced from ~1.8 – 1.9 GHz [16] to ~1.0 – 1.2 GHz, which is close to our experimental values. However, the simulation of ODMR resonances in the magnetic field-frequency map is very different from our result shown in Fig. 4(b). If the ClV defect had spin S=1, a typical monotonic splitting of the ODMR lines would be observed with increasing the magnetic field, without turning points of ODMR resonances at about 2, 8, and 13 G as observed in our experiments [Fig. 4(b)]. The features of the magnetic field-frequency map for our Cl-related defect are analogous to those of defects with spin S=3/2, similar to the Si vacancy in SiC [27], with turning points of ODMR lines occurring at magnetic fields corresponding to the level anti-crossing between the spin states. A conclusive identification of the electron spin of this defect and the symmetry of its configurations needs further investigation expected to be facilitated by single defect studies.

In summary, we have observed a Cl-related PL center in Cl-implanted 4H-SiC that emits at telecom bands. All four configurations observed for the center show stable emissions with narrow ZPLs. Our PL studies show that two centers emitting light at the C-band, Cl1 and Cl2, are characterized by narrow ZPLs and sharp LVMs with an estimated DW factor at least in the range of 22-25%. ODMR confirms that the defect is spin active with ZFSs in the range of 1.0 – 1.4 GHz, can be detected at room temperature, and shows a relatively high contrast (>5% at off-resonant 1525-nm excitation). Both PL and ODMR can be excited with laser energies close to resonant excitation without quenching eliminating the need of a repump laser. With emissions at the C-band and favorable optical and spin properties, this defect can be highly promising for developing quantum networks and remote quantum sensing. The experimental data concerning the DW factor and the ODMR dependence on external magnetic field does not agree well with the existing theoretical predictions [16,18] and further theoretical and experimental work is needed to find the spin magnitude and the microscopic structure of the observed Cl defect.

Support from the Knut and Alice Wallenberg Foundation (KAW 2018.0071) and the European Union (EU) under Horizon Europe for the projects QRC-4-ESP (Grant No. 101129663) and QUEST (Grant No. 101156088) is acknowledged. N.T.S. and I.G.I. acknowledge support from the EU project QuSPARC (Grant No. 101186889), Vinnova (Grant No. 2024-00461, 2025-03848, and QUASIC within QSIP, Grant No. 2024-03597). I.G.I., V.J. and N.T.S. acknowledge support from EU via the Swedish Agency for Economic and Regional Growth (project nr. 20370271). J.U.-H. acknowledges support from the EU project SPINUS (Grant No. 101135699) and the Swedish Research Council (VR) grant No. 2020-05444. I.G.I. acknowledges VR grant (VR 2025-06408). D.S. acknowledges support from AFM, Linköping University (CeNano grant 2021). I.A.A. is grateful for the support by the Swedish Government Strategic Research Area in Materials Science on Functional Materials at Linköping University (Faculty Grant SFOMat-LiU No. 2009 00971). L.V. acknowledges support from the Research Council of Norway through projects QuTe (no. 325573) and NorFab (no. 295864).

*tien.son.nguyen@liu.se

Table I: Line positions in nm (meV) and polarization of the ZPLs and LVMs of the four configurations of Cl-related center in 4H-SiC observed in this work. The energies and assignments of the four configurations of the Cl-vacancy defect predicted by theory [11] and observed by experiments in Ref. [12] are given for comparison.

| Line assignment | Wavelength nm (meV) | PL polarization | ZPL (nm) Theory [16] | ZPL (nm) Experiment [17] |
|---|---|---|---|---|
| Cl4 | 1351.1 (917.4) | $E_{PL} \perp c$ | ClV1: 1330 | ClV1: 1350 |
| Cl3 | 1396.7(887.4) | $E_{PL} \| c$ | ClV2: 1440 | ClV2: 1472 |
| Cl2' (ES of Cl2) | 1520.9 (815.0) | $E_{PL} \perp c$ | | |
| Cl1' (ES of Cl1) | 1526.7 (811.9) | $E_{PL} \perp c$ | | |
| Cl2 | 1533.3 (808.3) | $E_{PL} \perp c$ | ClV3: 1490 | ClV3: 1532 |
| Cl1 | 1539.2 (805.3) | $E_{PL} \perp c$ | ClV4: 1590 | ClV3: 1538 |
| LVM of Cl2 (LVM2) | 1557.4 (795.9) | $E_{PL} \perp c$ | | ClV4: 1556 |
| LVM of Cl1 (LVM1) | 1561.4 (793.8) | $E_{PL} \perp c$ | | ClV4: 1561 |
| Overtone of LVM1 and LVM2 | 1579.4 (784.8) | $E_{PL} \perp c$ | | |